# MULTISCALE MODELING OF THERMOMECHANICAL BEHAVIOUR OF THREE-PHASE NANOCOMPOSITE

Shailesh I. Kundalwal[a], and Shaker A. Meguid

*Mechanics and Aerospace Design Laboratory, Department of Mechanical and Industrial Engineering, University of Toronto, Toronto, Ontario, Canada*

*Summary* In this work, we developed an improved shear lag model to investigate the load transfer characteristics of three-phase nanocomposite which is reinforced with microscale fibers augmented with carbon nanotubes on their circumferential surfaces. The shear lag model accounts for (i) radial and axial deformations of different transversely isotropic constituents, (ii) thermomechanical loads on the representative volume element (RVE), and (iii) staggering effect of adjacent RVEs. The results from the current newly developed shear lag model are validated with the finite element simulations and found to be in good agreement. Our study reveals that the reduction in the maximum value of the axial stress in the fiber and the interfacial shear stress along its length become more pronounced in the presence of applied thermomechanical loads on the staggered RVEs. The existence of shear tractions along the RVE length plays a significant role in the load transfer characteristics and cannot be ignored.

## INTRODUCTION

Carbon nanotubes (CNTs) have emerged as ideal candidates for multifarious nanotechnology applications. This is due to their remarkable thermoelastic and physical properties. The quest for utilizing the exceptional thermoelastic properties of CNTs has led to the development of two-phase CNT-reinforced nanocomposites [1-4]. However, the addition of CNTs in polymer matrix does not always result in improved properties. Several important factors, such as defects, agglomeration and interface condition also play significant roles [1]. These difficulties can be alleviated using CNTs as secondary reinforcements in three-phase nanocomposites [5]. In this case, CNTs are grown on the circumference of the microscale fiber reinforcement [see Fig. 1]. The load transfer from the surrounding matrix to the fiber is one of the fundamental micromechanical processes determining the composite strength. However, existing shear lag studies do not account for the possible staggering effects of the adjacent RVEs [see Fig. 2] and cannot provide accurate description of the load transfer characteristics of this class of nanocomposites. The schematic diagram shown in Fig. 3 represents the three-phase nanocomposite made of a fiber, an intermediate CNT-reinforced composite (hereinafter the 'interlayer'), and a polymer matrix. In order to analyze the complex load transfer characteristics of such three-phase nanocomposite, a more accurate and consistent shear lag model that considers the staggering effect and the three-dimensional thermomechanical nature of the applied loads must be developed. This is indeed the motivation behind the current study.

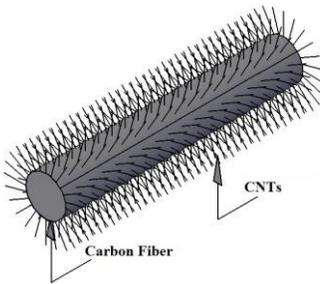 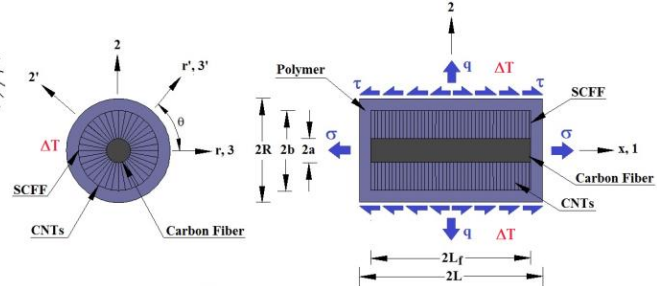 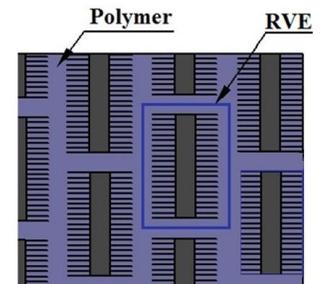

**Fig. 1.** Fiber grown with CNTs.    **Fig. 2.** RVE of three-phase composite.    **Fig. 3.** In-plane cross-section of composite lamina.

## ANALYTICAL AND FINITE ELEMENT SHEAR LAG DEVELOPMENT

Mori Tanaka model is used to determine the effective thermoelastic properties of the interlayer which are required as input to the shear lag model development; more details can be found in Ref. [6]. Unlike existing techniques, both the fiber and the interlayer are assumed to be transversely isotropic and undergo radial as well as axial deformations. In our model, the application of the shear stress ($\tau$) along the length of the RVE at $r = R$ accounts for the staggering of adjacent RVEs, whereas the applied radial load ($q$) on the RVE accounts for the lateral extensional interaction between the adjacent RVEs. Such radial/residual stresses may arise due to the manufacturing process of the short fiber composites. The cylindrical coordinate system ($r$–$\theta$–$x$) is considered in such a way that the axis of the RVE coincides with the x–axis, while CNTs are aligned along the r–direction. Note that the authors extended their earlier shear lag work [6] to incorporate the thermal loading and the radial as well as the axial deformations of transversely isotropic fiber and interlayer. The governing equations for an axisymmetric RVE problem in terms of cylindrical coordinates ($r$, $\theta$ and $x$) are given by

$$\frac{\partial \sigma_r^k}{\partial r} + \frac{\partial \sigma_{xr}^k}{\partial x} + \frac{\sigma_r^k - \sigma_\theta^k}{r} = 0 \quad \text{and} \quad \frac{\partial \sigma_x^k}{\partial x} + \frac{1}{r}\frac{\partial (r\sigma_{xr}^k)}{\partial r} = 0; \quad k = f, i \text{ and } m \qquad (1)$$

while the relevant constitutive relations are

$$\sigma_x^k = C_{11}^k \epsilon_x^k + C_{12}^k \epsilon_\theta^k + C_{13}^k \epsilon_r^k - \lambda_{11}^k \Delta T, \qquad \sigma_r^k = C_{13}^k \epsilon_x^k + C_{23}^k \epsilon_\theta^k + C_{33}^k \epsilon_r^k - \lambda_{22}^k \Delta T \text{ and } \sigma_{xr}^k = C_{66}^k \epsilon_{xr}^k \qquad (2)$$

---

[a] Corresponding author. Email: shailesh@mie.utoronto.ca

In Eqs. (1) and (2), the superscripts f, i and m denote the carbon fiber, the interlayer and the matrix, respectively. Where $\lambda_{11}^i$ and $\lambda_{22}^i$ are the axial and transverse thermal stiffness coefficients, respectively, and all other symbols have usual meaning. Subsequently, the following governing equations for the average axial stress in the fiber ($\sigma_x^f$) and the fiber-interlayer interfacial shear stress ($\tau_1$) are obtained using the boundary and loading conditions demonstrated in Fig. 2.

$$\sigma_x^f = A_1 \sinh(\alpha x) + A_2 \cosh(\alpha x) + A_3 \sinh(\beta x) + A_4 \cosh(\beta x) + (A_5/A_9)\sigma - (A_6/A_9)q - (A_7/A_9)\tau - (A_8/A_9)\Delta T \qquad (3)$$

$$\tau_1 = -\frac{a}{2}[A_{10}\alpha\cosh(\alpha x) + A_{11}\alpha\sinh(\alpha x) + A_{12}\beta\cosh(\beta x) + A_{13}\beta\sinh(\beta x)] \qquad (4)$$

The geometrical parameters of the RVE are taken as 2a = 10 μm, $L_f/a = 20$, $L/L_f = 1.1$, and $R/b = 1.1$; the volume fractions of carbon fiber and CNTs fixed to 0.4 and 0.019, respectively. For the thermoelastic properties of constituents, the readers are referred to Ref. [5]. To validate the analytical shear lag model, we developed three-phase finite element (FE) shear lag models using the commercial software ANSYS 14.0. With assumed hexagonal fiber packing array, a rectangular periodic RVE was defined. When the value of $\Delta T = 300$ K, it may be observed from Figs. 4(a) and 4(b) that if the staggering effect of the adjacent RVEs is ignored (i.e, $\tau = 0$), the analytical shear lag model over-estimates the values of $\sigma_x^f$ and $\tau_1$. It may also be observed that the good agreement between the two sets of results have been obtained and thus verifying the reliability of the analytical shear lag model incorporating the staggering effect and the thermal load. The variations of $\sigma_x^f$ and $\tau_1$ are presented in Figs. 5(a) and 5(b), respectively, for different values of applied thermal loads on the RVE. It may be observed from Fig. 5(a) that the maximum value of $\sigma_x^f$ is significantly decreased with the decrease in the magnitude of temperature change. This is attributed to the fact that the effective thermoelastic properties of the interlayer surrounding the fiber are improved with the decrease in the magnitude of temperature variation. On the other hand, the magnitude of temperature change does not much influence the value of $\tau_1$. Figures 4 and 5 reveal that the values of $\sigma_x^f$ remains uniform over 90% length of the fiber from its center while it decreases sharply near the end of the fiber. On the other hand, the values of $\tau_1$ reaches their maximum values near the ends of the fiber and become zero at $x = \pm L_f$.

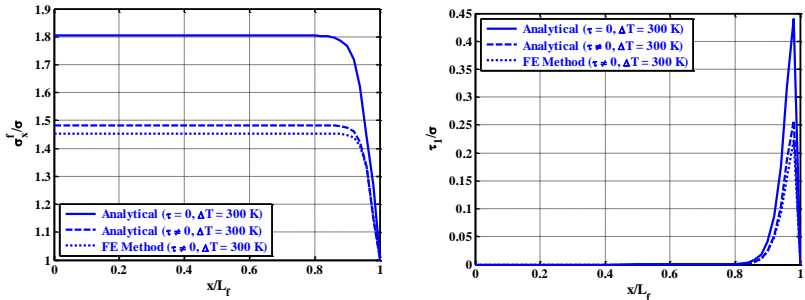

**Fig. 4.** Variation of (a) $\sigma_x^f$, and (b) $\tau_1$ along the fiber length.

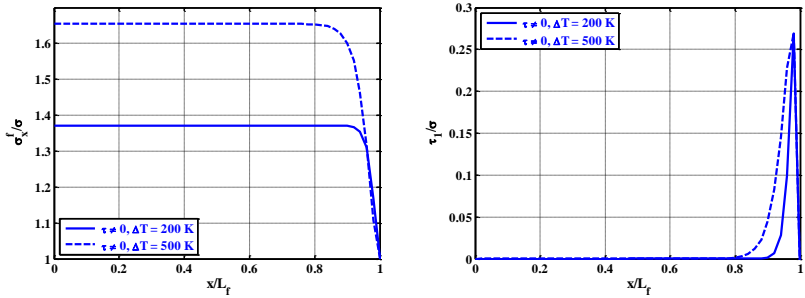

**Fig. 5.** Effect of temperature on the values of (a) $\sigma_x^f$, and (b) $\tau_1$.

## CONCLUSIONS

The improved shear lag model described herein provides an explanation to the thermomechanical load transfer mechanisms in three-phase nanocomposite considering transversely isotropic constituents and staggering effect which have not been previously considered elsewhere. Our study reveals that (i) the thermal loads significantly affect the average axial stress transferred to the microscale fiber, (ii) the existence of shear tractions along the length of the RVE play a crucial role in the load transfer characteristics of nanocomposite, and (iii) CNTs allow us to exploit their remarkable thermoelastic properties to improve the thermomechanical behavior of three-phase nanocomposite structures, enabling additional functionalities not available otherwise at the microscale. We will attempt to further this research through the investigation of effect of CNT orientations and debonding of microscale fiber.